\documentclass[11pt]{article}
\usepackage{moriond,epsfig,colordvi}

\def\Journal#1#2#3#4{{#1} {\bf #2}, #3 (#4)}

\def\NIM{\em Nucl. Instrum. Methods}

\def\PLB{{\em Phys. Lett.}  B}
\def\PRL{\em Phys. Rev. Lett.}
\def\PRD{{\em Phys. Rev.} D}

\def\EPJ{{\em Eur. Phys. J. direct}}

\def\be{\begin{equation}}
\def\ee{\end{equation}}
\def\bea{\begin{eqnarray}}
\def\eea{\end{eqnarray}}

\def\Ampl{{\cal A}}
\let \bar=\overline
\let \to=\rightarrow

\newcommand{\ssc}{\scriptscriptstyle}

\newcommand{\z}{_{\circ}}

\let\mathrm=\rm

\newcommand{\piz}{\pi^{\circ}}
\newcommand{\pip}{\pi^+}
\newcommand{\pim}{\pi^-}

\newcommand{\Km}{K^-}

\begin{document}
\newlength{\figwidth}
\newlength{\figheight}
\vspace*{4cm}
\title{LIGHT MESONS FROM CHARM \\ MESON DECAYS IN FNAL E791}

\author{ B. T. MEADOWS
\footnote{Representing the E791 collaboration.}
}
\address{Department of Physics \\ University of Cincinnati
\footnote{Supported by NSF grant 9901568.}
\\
Cincinnati, OH, USA}

\maketitle\abstracts{
An analysis of the Dalitz plots for the decays $D^+\to\Km\pip\pip$ and
$D^+\to\pim\pip\pip$ indicates that structure with significant
phase variation is required in the $s$-wave $\Km\pip$ and $\pim\pip$
systems in the region below 1 GeV/c$^2$ effective mass.  A constant,
non resonant ``contact" term, together with a simple $s$-wave Breit
Wigner amplitude with
$M_{\z}=797\pm 19\pm 43, \Gamma_{\z}= 175\pm 12\pm 12$~MeV/c$^2$ for the
$\Km\pip$ system and
$M_{\z}=478^{+24}_{-23}\pm 17,~\Gamma_{\z}=324^{+42}_{-40}\pm 21$~MeV/c$^2$
for the $\pim\pip$ system provide good fits to the data in this region.}

\section{Introduction}
%
%
Decays of $D$ mesons to three pseudo scalar mesons can produce intermediate,
two meson systems with natural parity.  Kinematics and angular momentum
barrier factors generally favour scalar ($J^P=0^+$) over vector $1^-$ or
tensor $2^+$ systems.  Analysis of these decays can therefore provide new
information on the poorly understood scalar meson system.

In most instances, $D$ decays appear to proceed through quasi two body
channels.  In the cases of $D^+$ decays
\footnote{Except where indicated otherwise, charge conjugate systems are
implied in this paper.}
to $\pim\pip\pip$ and to $\Km\pip\pip$ however, intermediate states
found in the Particle Data Group listings \cite{pdg} (PDG) have been
found to contribute only partially to the decays, with three body modes
apparently dominant \cite{e691,e687:kpipi}.  We report here on analyses
\cite{e791:d3pi,e791:kpipi} of these decays using the largest samples so
far examined.

\section{Fermilab Experiment E791}

In this experiment \cite{e791:exp} $2\times 10^{10}$ charm enhanced
interactions between 500 GeV/c $\pim$ mesons and thin foil targets
(four $C$, one $Pt$) are recorded \cite{e791:odf}.  Foils are spaced so
that decays of $D$ mesons produced in them occur predominantly
between them.  Silicon strip detectors measure
$D$ decay lengths (mean $\sim$5~mm) with a typical precision of
350 $\mu$m.
Charged particle momenta are measured with two dipole magnets and
drift chambers and two threshold gas Cherenkov detectors provide
$K/\pi/p$ separation in the 6-60~GeV/c momentum range.

%
Events with one negative and two positive charged tracks making a
good vertex, well separated from any foil and from
the primary interaction point, are selected for further analysis.
The resultant momentum is required to point back to within 
40~$\mu$m of the primary vertex. For the $\Km\pip\pip$ sample the 
$\Km$ is required to have momentum $6$~GeV/c$<|\vec p_K|<40$~GeV/c
and to be identified by the Cherenkov detectors.
Other cuts are also made to reduce contamination from
tracks that are really associated with the primary vertex.
With these selection criteria, the major sources of background are
mostly misidentified (or incomplete) charm decays, with a small,
``combinatorial" contribution from false secondary vertices.  
Yields of 15,090 $\Km\pip\pip$
(6\% background) and 1,686 $\pim\pip\pip$ (28\% background) remain
in the signal regions.

\section{Isobar Analysis of $D$ Decay to Three Pseudo-Scalars}
\label{sec:isobar}

Decays of $D$ mesons to three pseudo scalars $i,~j,~k$
can be described approximately as a sum of isobar amplitudes, each 
corresponding to a quasi two body decay $D\to R(\to i~j)~k$.
Each amplitude must satisfy Lorentz invariance and conserve total spin 
and has the form
 \begin{eqnarray*}
   \Ampl_{\ssc R}(s_{ij}, s_{ik}) &=&
          F_{\ssc D}(q,r_{\ssc D})
          F_{\ssc R}(p,r_{\ssc R})
          \times BW_{\ssc R}(s_{ij})
          \times~(-2)^J |\vec p|^J |\vec q|^J P_J(\cos\theta)
 \end{eqnarray*}
where $\vec p, \vec q$ are momenta of $i$ and $k$ respectively in 
the $(ij)$ rest frame and $\cos\theta=\hat p\cdot\hat q$.
Form factors $F_{\ssc D}$ ($F_{\ssc R}$) for $D$ ($R$) are parametrized 
in terms of  effective radii $R_{\ssc D}$ ($R_{\ssc R}$).  A Breit Wigner 
propagator
$BW_{\ssc R}=\left[s_{\ssc R} - s_{ij} -
               i\sqrt{s_{\ssc R}}\;\Gamma(s_{ij}) \right]^{-1}$
describes the resonance $R$ with spin $J$, mass $\sqrt{s_{\ssc R}}$.
Suffix $R$ denotes a quantity evaluated at $s_{ij}=s_{\ssc R}$.

A decay is kinematically specified by coordinates $(s_{ij}, s_{ik})$
on a Dalitz plot and by the
three body mass $M_{ijk}$ whose distribution about the $D$ mass $m_D$
due to resolution is $G(M_{ijk})$, determined from the data.
The signal distribution of decays is
  \[ {\cal P}_{\ssc S}(s_{ij}, s_{ik}, M_{ijk}) ~=~G(M_{ijk})\times
     \left|\, a_{\ssc NR}e^{i\delta_{\ssc NR}}~+~\sum_{\ssc R} a_{\ssc R}
     e^{i\delta_{\ssc R}}\Ampl_{\ssc R}(s_{ij}, s_{ik})\, \right|^2 \]
in which three body decays (NR) are described by a constant,
``contact" amplitude $a_{\ssc NR}e^{i\delta_{\ssc NR}}$.

The distribution ${\cal P}_{\ssc B}(s_{ij}, s_{ik}, M_{ijk})$ of
backgrounds is modelled using events from side bands in $M_{ijk}$ for
combinatorial, and from MC simulations for broken charm sources.  The
efficiency $\epsilon(s_{ij},s_{ik})$ for observing decays is
determined from MC simulations.

Complex coefficients $ae^{i\delta}$ and resonance parameters are 
determined from fits to a linear combination of 
$\epsilon{\cal P}_{\ssc S}$ and ${\cal P}_{\ssc B}$ in the
correct proportions.  The fraction of $D$ decay to an isobar $R$ is
$f_{\ssc R} = G(M_{ijk})\int |{\cal A}_{\ssc R}|^2 ds_{ij}ds_{ik} /
              \int{\cal P}_{\ssc S} ds_{ij}ds_{ik}$
evaluated at $M_{ijk}=m_{\ssc D}$.

\section{The $\Km\pip\pip$ Channel.}

\begin{figure}[ht]
 \begin{minipage}[t]{0.48\textwidth}
  \setlength{\figwidth}{0.90\textwidth}
  \setlength{\figheight}{0.80\textwidth}
  \centerline{%
  \epsfig{file=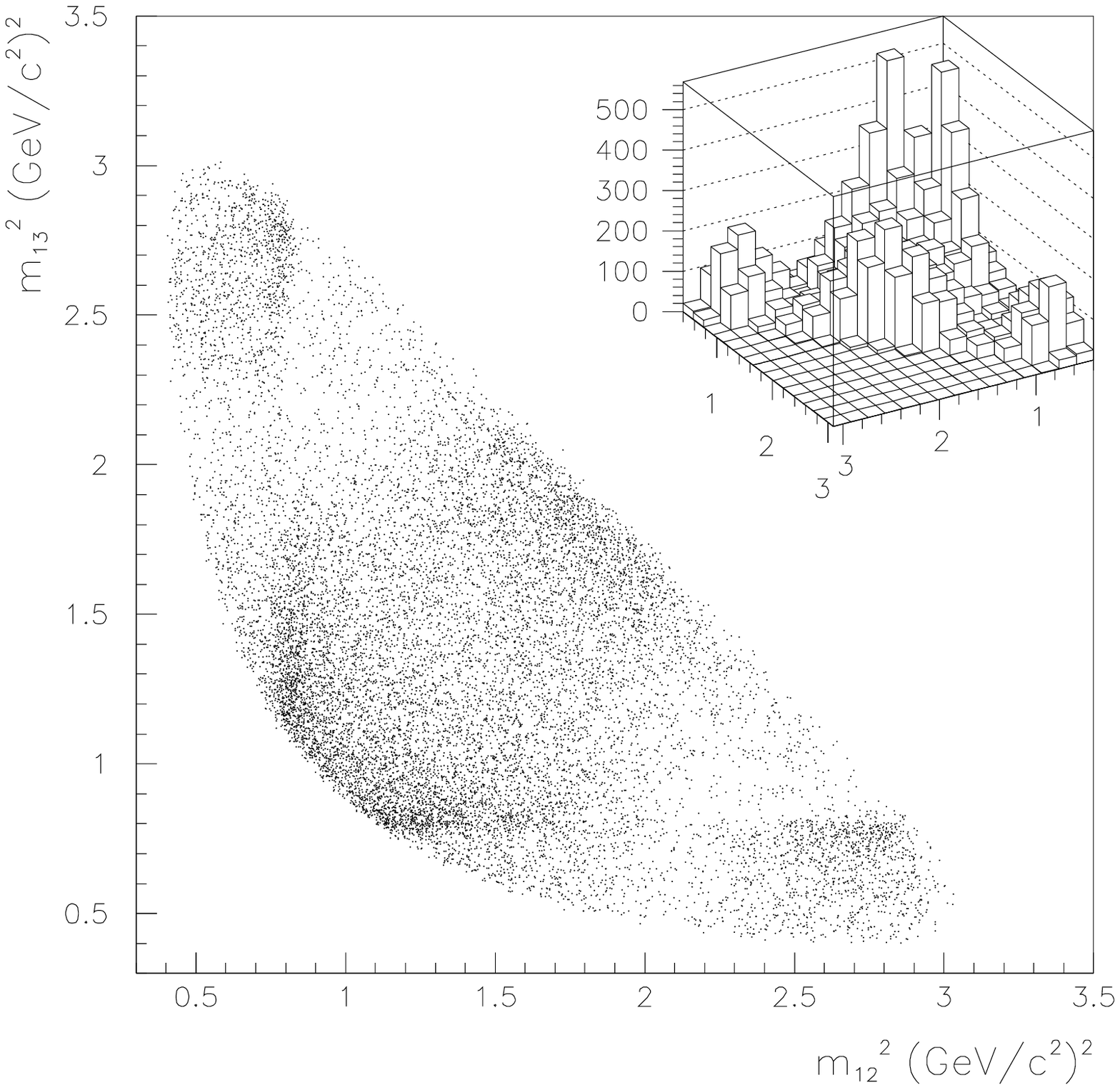,
       height=0.95\figheight,
       width=0.95\figwidth,angle=0}}
  \large
  \vskip-0.9\figheight\hskip0.30\figwidth{\bf (a)}
  \par\vspace{0pt}
 \end{minipage}
 \hspace{0.00\textwidth}
 \begin{minipage}[t]{0.48\textwidth}
  \setlength{\figwidth}{0.90\textwidth}
  \setlength{\figheight}{0.80\textwidth}
  \centerline{%
  \epsfig{file=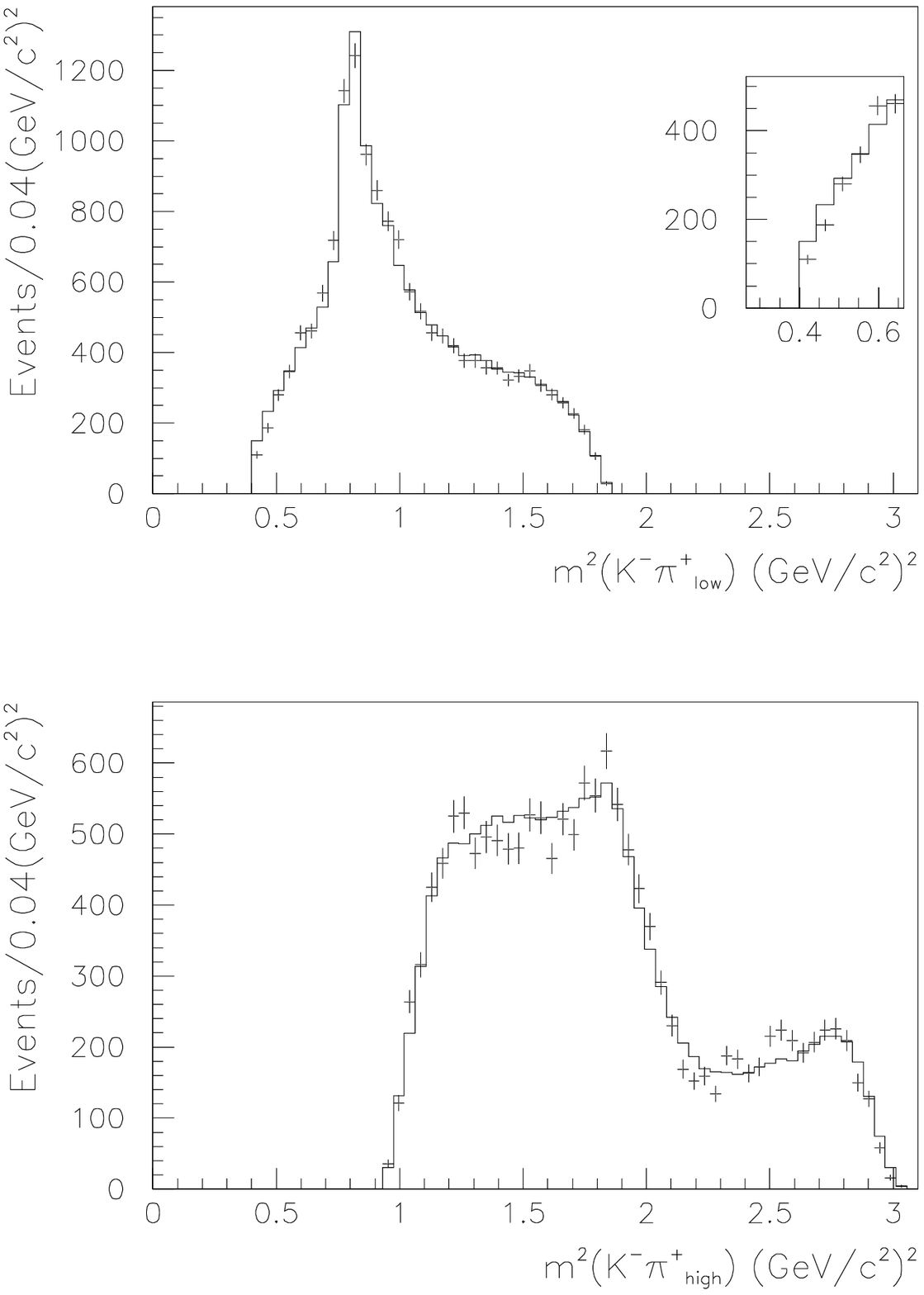,
          width=0.95\figwidth,
          height=\figheight,
          angle=0}}
  \large
  \vskip-0.30\figheight\hskip.86\figwidth
  \Red{$\mathbf\downarrow$}\hskip-5.5pt
  \Red{$\mathbf\downarrow$}
  \vskip-0.58\figheight\hskip0.30\figwidth
  \Red{$\mathbf\searrow$}\hskip-11.5pt
  \Red{$\mathbf\searrow$}
  \vskip-0.22\figheight\hskip0.87\figwidth
  \Red{$\mathbf\searrow$}\hskip-11.5pt
  \Red{$\mathbf\searrow$}
  \vskip-.11\figheight\hskip0.28\figwidth{\bf (b)}
  \vskip0.49\figheight\hskip0.28\figwidth{\bf (c)}
  \par\vspace{0pt}
 \end{minipage}
 \caption{(a) Dalitz plot for $D^+\to\Km\pip\pip$ decays.
  Projections for (b) smaller and (c) larger
  $K\pi$ effective mass for data (crosses)
  and fit (line) to model I described in
  the text.  Arrows indicate regions where the fit is worst.
 \label{fig:kpipi}}
 \vskip-6pt
\end{figure}
The Dalitz plot in figure \ref{fig:kpipi}(a) shows clear bands in
the $\bar{K}_{1}(890)$ regions with an asymmetric structure
indicative of interference with a significant, underlying
$s$-wave.  An accumulation of events near 2~(GeV/c$^2)^2$
indicates contributions from $\bar{K}_{\z}(1430)$ or
$\bar{K}_{2}(1430)$.  A fit is made including isobar amplitudes 
for these resonances and the $\bar{K}_{1}(1688)$, each with
mass and width fixed at their PDG values.  Results
are shown in table \ref{tab:d3} (model I) and compared with the
$\Km\pip$ mass projections in figures \ref{fig:kpipi}(b) and (c).
As in previous analyses \cite{e691,e687:kpipi} using this model, a
large NR component is required ($\sim 90$\%) and the sum of all
contributions is $\sim 140$\% indicating much destructive
interference.
\begin{table}[ht]
\caption{Resonant fractions $f$ for $D^+\to\Km\pip\pip$,
and $\pim\pip\pip$.  Models I-IV are described in the text.
\label{tab:d3}}
\vspace{0.4cm}
\begin{center}
\begin{tabular}{|ccccccc|}
\hline
\\[-12pt]
$D^+\to$
& $\Km\pip\pip$   & $\kappa\pi$       & $K_1(890)\pi$
& $K_{0}(1430)\pi$
                  & $K_2(1430)\pi$    & $K_1(1688)\pi$
\\[1pt] \hline \\[-12pt]
I
& $90.0\pm 2.6$   & -                 & $13.8\pm 0.5$ 
& $30.6\pm 1.6$   & $0.4\pm 0.1$      & $3.2\pm 0.3$
\\[0pt]
II
& $13.0\pm 5.8$   & $47.8\pm 12.1$    & $12.3\pm 1.0$
& $12.5\pm 1.4$   & $0.5\pm 0.1$      & $2.5\pm 0.7$
\\[-3pt]
& $\pm 4.4$       & $\pm 5.3$         & $\pm 0.9$
& $\pm 0.5$       & $\pm 0.2$         & $\pm 0.3$
\\[1pt] \hline \\[-12pt]
$D^+\to$
& $\pim\pip\pip$   & $\sigma\pi$      & $\rho(770)\pi$
& $f_{0}(980)\pi$ & $f_2(1270)\pi$   & $\rho(1450)\pi$
\\[1pt] \hline \\[-12pt]
III
& $38.6\pm 1.4$    & -                & $20.8\pm 2.3$ 
& $ 7.4\pm 4.3$    & $6.3\pm 3.3$     & $22.6\pm 2.1$
\\[0pt]
IV
& $ 7.8\pm 6.0$    & $46.3\pm 9.0$    & $33.6\pm 3.2$
& $ 6.2\pm 1.3$    & $19.4\pm 2.5$    & $ 0.7\pm 0.7$
\\[-3pt]
& $\pm 2.7$        & $\pm 2.1$        & $\pm 2.2$
& $\pm 0.4$        & $\pm 0.4$        & $\pm 0.3$
\\[-1pt] \hline
\end{tabular}

\vskip-6pt
\end{center}
\end{table}
The fit is poor with a $\chi^2$, found by comparing predictions of
model I with data in equal square bins in the Dalitz plot, of 167
for 63 degrees of freedom $\nu$.  Major discrepancies are at
low $\Km\pip$ masses, and the corresponding high mass reflection, and
can be seen in earlier data for both this channel \cite{e687:kpipi}
and for $D^{\z}\to\Km\pip\piz$ \cite{cleokpipi}.  However the large
E791 sample underlines their significance, and demands that a better
model be found.

First, the radii $R_{\ssc D}$, $R_{\ssc R}$ and the mass and width of
$K_{0}(1430)$ are allowed to float \cite{wmd}.  This produces little
improvement.  Next, a further scalar amplitude ``$\kappa"$ having a
Breit Wigner form similar to $K_{0}(1430)$ is added (model II in
Table \ref{tab:d3}).  This does provide an excellent description of
the data with $\chi^2/\nu=46/63$.  The $\kappa$ mode is dominant while
the NR component is consistent with zero.  The sum of fractions is
$\sim 86$\%.  The $\kappa$ mass and width, determined from the fit,
are $M_{\z}=797\pm 19\pm 43$ and $\Gamma_{\z}= 175\pm 12\pm 12$~MeV/c$^2$.
In this model, however, the $K_{0}(1430)$ mass and width are found to be
$1459\pm 7\pm 12$~MeV/c$^2$ and $175\pm 12\pm 12$~MeV/c$^2$ respectively,
differing somewhat from the values from LASS \cite{wmd}.

The $s$-wave BW form used for the $\kappa$ in model II appears to be
important.  Variations in this, or in form factors for $\kappa$ and NR
terms, are found to affect principally the amounts of $\kappa$ and NR,
but to have relatively small effects on $\kappa$ mass and width
parameters.  Replacing the complex BW in model II by its absolute
magnitude (a ``$\kappa$" mass enhancement, but no phase variation)
results in unphysically large $\kappa$ and NR fractions.  A vector form for
$\kappa$ converges with mass $\sim 1100$~MeV/c$^2$ and negligible fraction.
A tensor $\kappa$ fit fails to converge.

\section{$D^+\to\pim\pip\pip$ Decays.}
\begin{figure}[ht]
 \begin{minipage}[b]{0.38\textwidth}
  \setlength{\figwidth}{0.75\textwidth}
  \setlength{\figheight}{0.75\textwidth}
  \centerline{%
  \epsfig{file=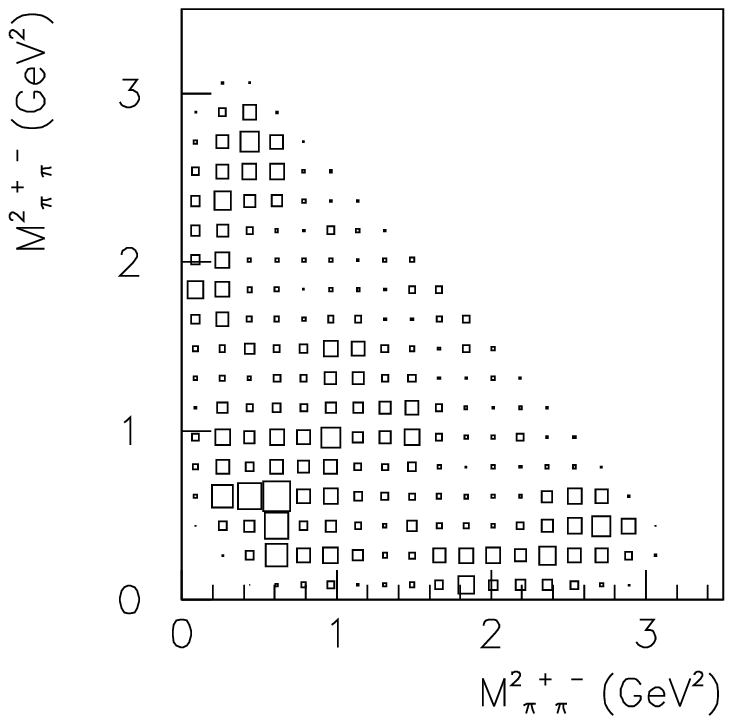,
          width=\figwidth,
          height=\figheight,
          angle=0}}
  \large
  \vskip-0.85\figheight\hskip0.85\figwidth{\bf (a)}
  \vskip0.63\figheight
  \par\vspace{0pt}
 \end{minipage}
 \hspace{0.00\textwidth}
 \begin{minipage}[b]{0.57\textwidth}
  \setlength{\figwidth}{0.75\textwidth}
  \setlength{\figheight}{0.45\textwidth}
  \centerline{%
  \epsfig{file=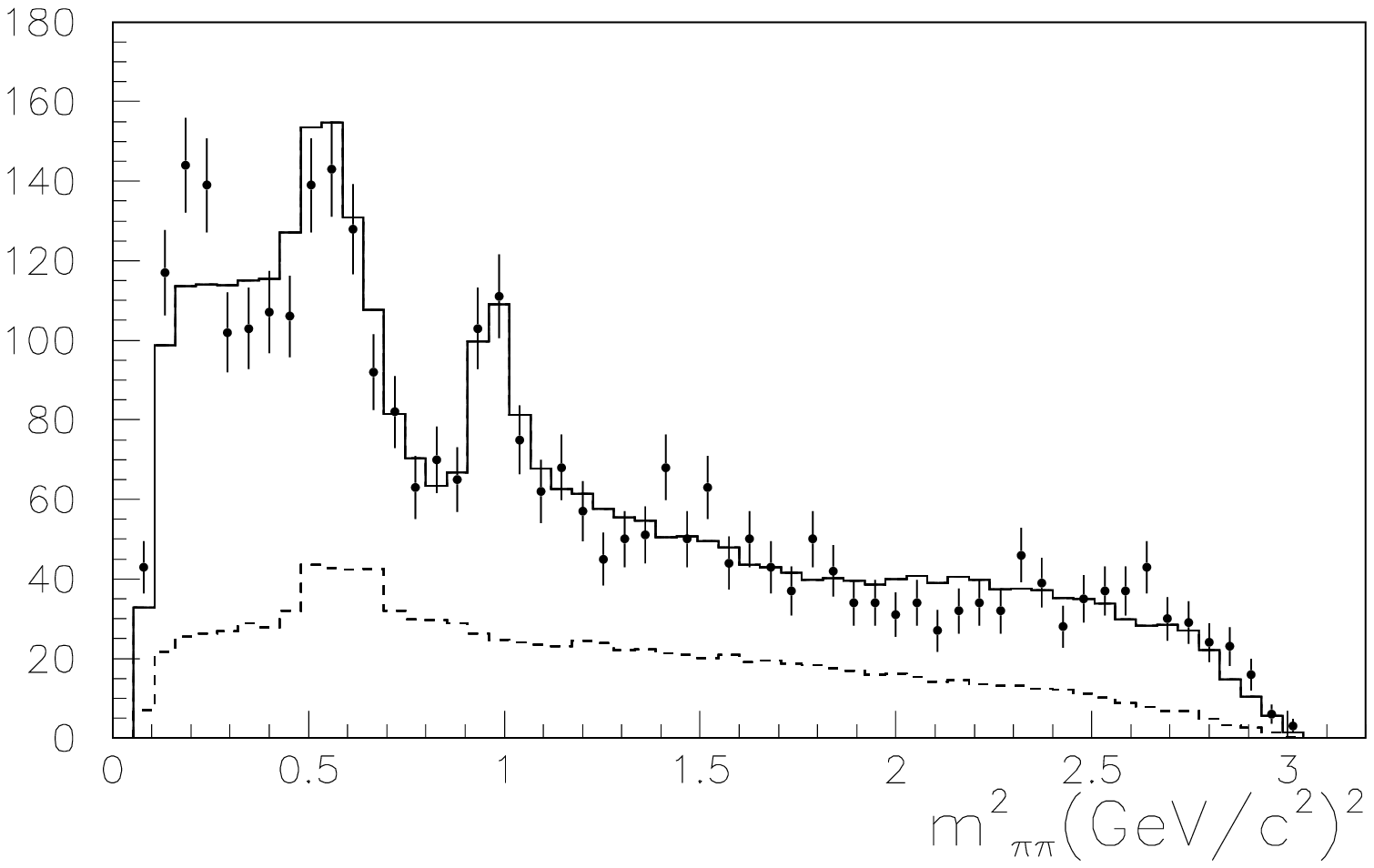,
          width=\figwidth,
          height=\figheight,
          angle=0}}
  \large
  \vskip-.85\figheight\hskip1.00\figwidth{\bf (b)}
  \vskip0.72\figheight
  \par\vspace{0pt}
 \end{minipage}
 \caption{(a) Dalitz plot for $D^+\to\pim\pip\pip$ decays.
  (b) Projection of $\pim\pip$ effective mass.  Data
  (points with errors) and predictions (solid line) from model III
  described in the text indicate poor agreement at low mass.
 \label{fig:3pi}}
\end{figure}

Figure \ref{fig:3pi}(a) is the Dalitz plot for these decays \cite{e791:d3pi}.
Table \ref{tab:d3} shows results for model III in which only resonances from
the PDG are included.  The fit is poor ($\chi^2/\nu=1.5$), especially in the
low mass $\pim\pip$ system as seen in figure \ref{fig:3pi}(b).  The
contribution from $\rho(1450)$ is larger than that from $\rho(770)$ -
a somewhat unlikely result.  As observed in the previous analysis
\cite{e687:d3pi} of this channel with a similar model, the NR mode dominates.

Adding a scalar amplitude ($\sigma$, similar to the $\kappa$ in
model II) results in a superior fit ($\chi^2/\nu=0.9$).  Results for this
(model IV) are in table \ref{tab:d3}.  Mass and width for $\sigma$ are found
to be $M_{\z}=478^{+24}_{-23}\pm 17$ and
$\Gamma_{\z}=324^{+42}_{-40}\pm 21$~MeV/c$^2$.
Both NR and $\rho(1450)\pi$ contributions are insignificant, and the
$\sigma\pi$ amplitude is dominant.
As for the $\kappa$, the $s$-wave BW form seems to be the best model
of those applied to these data.

\section{Conclusions.}

We conclude that the $s$-wave $\Km\pip$ system from decay of $D^+$
to $\Km\pip\pip$ has amplitude and phase variation consistent with a
sum of NR, $K_{0}(1430)$ and scalar $\kappa$ terms.  Similarly, the
$s$-wave $\pim\pip$ system from decay of $D^+\to\pim\pip\pip$
has amplitude and phase variation consistent with a sum of NR,
$f_{0}(980)$ and scalar $\sigma$ terms.  Though this hints at the
existence of scalar $\kappa$ and $\sigma$ resonances, use of Breit
Wigner functions to describe such broad states near threshold may not be
appropriate, and other interpretations of the data may be possible.

\end{document}